\documentclass[11pt]{article}
\usepackage{epsfig}
\usepackage{amsfonts}
\usepackage{amsmath,amssymb}
\usepackage{bbm,bm}
\usepackage{soul}
\usepackage{cite}
\usepackage{url}
\allowdisplaybreaks[1]

\usepackage{tikz}
\usetikzlibrary{shapes,arrows,shadows,automata,positioning}
\newdimen\nodeDist
\usepackage{comment}
\usepackage[normalem]{ulem} 

\usepackage[
  colorlinks=true,
  linkcolor={red!50!black},
  citecolor={blue!50!black},
  filecolor=black,
  urlcolor=black,
  breaklinks=true
  ]{hyperref}
\usepackage{orcidlink}

\colorlet{CustomBlue}{blue!50}
\definecolor{CustomRed}{RGB}{207,0,99}

\newcommand{\nn}{\nonumber \\}
\renewcommand{\vec}[1]{{\bf #1}}
\newcommand{\rmi}[1]{{\mbox{\scriptsize #1}}}
\newcommand{\rmii}[1]{{\mbox{\tiny\rm{#1}}}}
\newcommand{\clog}{\Big(c+\ln\Big(\frac{3T}{\Lamd}\Big)\Big)}
\newcommand{\bmu}{\bar{\mu}}
\newcommand{\Lamd}{\bmu_{\rmii{3d}}}
\newcommand{\LamD}{\bmu}
\newcommand{\LamDref}{\LamD_\text{ref}}
\newcommand{\LamDin}{\LamD_0}
\newcommand{\LamX}{\mathcal{N}}

\newcommand{\gammaE}{{\gamma_\rmii{E}}}
\newcommand{\Veff}{V_{\text{eff}}}
\newcommand{\Seff}{{S_{\text{eff}}}}

\newcommand{\Tc}{T_{\rm c}}
\newcommand{\Treh}{T_\text{reh}}

\newcommand{\Tp}{T_{\rm p}}
\newcommand{\vw}{v_w}

\newcommand{\gX}{g_\rmii{$X$}}
\newcommand{\gXt}{g_{\rmii{$X$},3}}

\newcommand{\mD}{m_\rmii{D}}

\newcommand{\mX}{m_\rmii{$X$}}
\newcommand{\mXt}{m_{\rmii{$X$},3}}
\newcommand{\MX}{M_\rmii{$X$}}

\newcommand{\Mpl}{M_\rmii{Pl}}

\newcommand{\fp}{f_{\rm p}}

\newcommand{\phibg}{\varphi}
\newcommand{\phib}{\phibg_b}
\newcommand{\phiField}{\phi}

\renewcommand{\det}{\mbox{det}}

\newcommand{\T}{\rmii{$T$}}
\newcommand{\Tstar}{T_\star}
\newcommand{\Hstar}{H_\star}

\newcommand\MSbar{$\overline{\rm MS}$}

\newcommand{\geff}{g_*}
\newcommand{\heff}{h_*}

\newcommand{\dd}{{\rm d}}

\usepackage{pifont}

\newcommand{\article}{paper}

\newcommand{\UOneX}{{\rm U}(1)_\rmii{$X$}}
\newcommand{\SUTwoX}{{\rm SU}(2)_\rmii{$X$}}
\newcommand{\UOneY}{{\rm U}(1)_\rmii{$Y$}}

\newcommand{\QX}{Q_\rmii{$X$}}

\newcommand{\vphi}{v_\phi}
\newcommand{\lh}{\lambda_h}
\newcommand{\lphi}{\lambda_\phiField}
\newcommand{\lp}{\lambda_p}
\newcommand{\NLOdet}{{\tt NLOdet}}
\newcommand{\Daisy}{{\tt daisy}}

\newcommand{\lphit}{\lambda_{\phi,3}}

\newcommand{\Lb}{L_b}
\newcommand{\Lf}{L_f}

\newcommand{\nB}{n_\rmii{B}}

\makeatletter \@addtoreset{equation}{section} \makeatother
\renewcommand{\theequation}{\arabic{section}.\arabic{equation}}
\makeatletter
\renewcommand\section{\@startsection{section}{1}{\z@}%
  {-5.5ex \@plus -1ex \@minus -.2ex}
  {2.3ex \@plus.2ex}%
  {\normalfont\large\bfseries}}
\renewcommand\subsection{\@startsection{subsection}{2}{\z@}%
  {-3.25ex\@plus -1ex \@minus -.2ex}%
  {1.5ex \@plus .2ex}%
  {\normalfont\normalsize\bfseries}}
\renewcommand\thesection{\@arabic\c@section}
\renewcommand\thesubsection{\thesection.\@arabic\c@subsection}
\renewcommand{\@seccntformat}[1]{%
  \csname the#1\endcsname.\hspace{1.0em}}
\makeatother

\begin{document}

\flushbottom

\begin{titlepage}

\begin{flushright}
KA-TP-17-2026
\end{flushright}
\begin{centering}

\vfill

{\Large{\bf
  Theoretical uncertainties in reconstructing model parameters\\
  with gravitational waves from supercooled phase transitions%
}}

\vspace{0.6cm}

\renewcommand{\thefootnote}{\fnsymbol{footnote}}
Maciej Kierkla%
\orcidlink{0000-0002-2785-5370}%
,$^{\rm a,}$%
\footnotemark[1]
Marek~Lewicki%
\orcidlink{0000-0002-8378-0107}%
,$^{\rm b,c,}$%
\footnotemark[2]
Philipp Schicho%
\orcidlink{0000-0001-5869-7611}%
,$^{\rm d,}$%
\footnotemark[3]
\\
Daniel~Schmitt%
\orcidlink{0000-0003-3369-2253}%
,$^{\rm e,f,}$%
\footnotemark[4]
and
Bogumi{\l}a {\'S}wie{\.z}ewska%
\orcidlink{0000-0003-0169-211X}%
\,$^{\rm b,}$%
\footnotemark[5]

\vspace{0.6cm}

$^\rmi{a}$%
{\em
Department of Physics and Astronomy, Uppsala University,\\
Regementsv\"agen~10, 752 37 Uppsala, Sweden\\}
\vspace{0.3cm}

$^\rmi{b}$%
{\em
Faculty of Physics, University of Warsaw,
Pasteura~5, 02-093 Warsaw, Poland\\}
\vspace{0.3cm}

$^\rmi{c}$%
{\em 
Astrocent, Nicolaus Copernicus Astronomical Center Polish Academy of Sciences,\\
ul.\ Rektorska~4, 00-614, Warsaw, Poland\\}
\vspace{0.3cm}

$^\rmi{d}$%
{\em
  D\'epartement de Physique Th\'eorique, Universit\'e de Gen\`eve,\\
  24 quai Ernest Ansermet, CH-1211 Gen\`eve 4, Switzerland\\}
\vspace{0.3cm}

$^\rmi{e}$%
{\em
Institute for Astroparticle Physics (IAP), Karlsruhe Institute of Technology (KIT),
Hermann-von-Helmholtz-Platz~1, 76344 Eggenstein-Leopoldshafen, Germany\\}
\vspace{0.3cm}

$^\rmi{f}$%
{\em
Institute for Theoretical Physics (ITP), Karlsruhe Institute of Technology (KIT),
Wolfgang-Gaede-Str.~1, 76131 Karlsruhe, Germany\\}

\vspace*{0.6cm}

\mbox{\bf Abstract}

\end{centering}

\vspace*{0.3cm}

\noindent
Future interferometers may detect a
gravitational-wave (GW) signal from
a cosmological first-order phase transition.
Reconstructing the underlying particle-physics model from such a signal
requires theoretical control over the map from microphysics to the spectrum.
For classically scale-invariant extensions of the Standard Model,
which generically predict strongly supercooled transitions and
strong GW signals, this map depends sensitively on
the treatment of quantum and thermal corrections to the nucleation rate.
Taking the classically conformal $\UOneX$ model as representative
of this class, we scan its parameter space and compare two resummation schemes.
The first is a high-temperature effective field theory,
matched at two-loop level and including next-to-leading-order corrections to
the bounce action, with the nucleation-rate prefactor given by
the full one-loop functional determinants.
The second is a commonly employed daisy-resummed effective potential,
with the prefactor estimated on dimensional grounds.
Reconstructing the fundamental model parameters through a Fisher-matrix analysis of
injected GW signals at LISA,
we find that the daisy-resummation scheme is strongly disfavored,
as its theoretical error dominates over
the reconstruction uncertainty.

\end{titlepage}

{\hypersetup{hidelinks}
\tableofcontents
}

\renewcommand{\thefootnote}{\fnsymbol{footnote}}
\footnotetext[1]{maciej.kierkla@physics.uu.se}
\footnotetext[2]{marek.lewicki@fuw.edu.pl}
\footnotetext[3]{philipp.schicho@unige.ch}
\footnotetext[4]{daniel.schmitt@kit.edu}
\footnotetext[5]{bogumila.swiezewska@fuw.edu.pl}
\clearpage

\renewcommand{\thefootnote}{\arabic{footnote}}
\setcounter{footnote}{0}

%
\section{Introduction}
\label{sec:intro}

The evidence for a stochastic gravitational wave (GW) background in
pulsar-timing-array data~\cite{NANOGrav:2023gor,EPTA:2023fyk,Reardon:2023gzh,Xu:2023wog}
which could still prove to be of primordial origin~\cite{NANOGrav:2023hvm,Figueroa:2023zhu,Ellis:2023oxs}
and the prospects of
upcoming space-based interferometers such as
LISA~\cite{Caprini:2015zlo,LISA:2017pwj,LISACosmologyWorkingGroup:2022jok,Caprini:2019egz} or
Taiji~\cite{Ruan:2018tsw}
have turned cosmological first-order phase transitions (PTs)
into a quantitative probe of physics beyond the Standard Model (SM).

A phase transition associated with electroweak symmetry breaking
is expected to have occurred in the early universe.
Within the pure SM this transition is
a smooth crossover~\cite{Kajantie:1996mn,Gurtler:1997hr,Csikor:1998eu},
so that no GW signal is produced.
Many viable extensions of the SM, however, feature a first-order phase transition
that releases energy and can source a stochastic GW background
within the LISA sensitivity band.
LISA data will thus provide an unprecedented source of information
about physics at fundamental scales.
The simplest criterion to assess
the compatibility of a model with the data is binary:
whether or not
the model predicts a strong enough GW signal.
A detected signal, however,
carries far more information.
Assuming a certain spectral template,
one can
directly reconstruct
the fundamental parameters of the underlying model~\cite{Caprini:2019pxz,Caprini:2024hue}.
This reconstruction faces several obstacles.
First, the expected signal is typically weak and can be
buried under astrophysical galactic and extragalactic foregrounds%
 \cite{%
 Robson:2018ifk,Cornish:2017vip,
 Boileau:2020rpg,Boileau:2021sni,
 Babak:2023lro,
 Pieroni:2020rob,Lewicki:2021kmu,Racco:2022bwj,
 Hindmarsh:2024ttn}.
Moreover, degeneracies among the thermodynamic parameters imply
that different parameter sets can yield
the same spectrum~\cite{%
  Gowling:2021gcy,Giese:2021dnw,Boileau:2022ter,Gowling:2022pzb,Caprini:2024hue}.

Another obstacle in this program is the reliability of the
thermodynamic input.
Predictions of the transition strength $\alpha$,
the inverse duration $\beta/H$, and
the percolation temperature $\Tp$
depend sensitively on how thermal fluctuations are resummed in the
effective action and the bubble nucleation
rate~\cite{Croon:2020cgk,Gould:2023ovu}.
Different levels of computational diligence,
ranging from four-dimensional (4D) daisy-resummed potentials
to next-to-leading order (NLO)
within the dimensionally reduced three-dimensional (3D)
effective field theory (EFT)~\cite{%
  Schicho:2021gca,Niemi:2021qvp,Schicho:2022wty},
and different approaches to evaluating the nucleation-rate prefactor
can yield quantitatively, and sometimes qualitatively, different
predictions.
As emphasized in~\cite{Caprini:2024hue}
and in the analysis of the real-singlet extension~\cite{Lewicki:2024xan},
these theoretical uncertainties propagate into the
\emph{reconstructed} model parameter space and can dominate over the
experimental uncertainty of a LISA measurement.

In this \article{} we focus on the classically conformal
$\UOneX$ extension of the SM~\cite{Iso:2009ss},
a minimal gauge-Higgs model in which electroweak symmetry breaking is
triggered radiatively \`a la Coleman-Weinberg~\cite{Coleman:1973jx}. 
In models with strong dynamics or extra dimensions, see e.g.~\cite{%
  Randall:2006py,Konstandin:2010cd,Konstandin:2011dr,vonHarling:2017yew,
  Bruggisser:2018mrt,Kubo:2016kpb,Baldes:2021aph},
and in models with
perturbative classically scale-invariant potentials~\cite{%
  Hambye:2013dgv,Jaeckel:2016jlh,Hashino:2016rvx,Jinno:2016knw,Marzola:2017jzl},
\emph{strongly supercooled} first-order PTs are generically predicted,
which are among the most promising GW sources.
Since these signals are generically strong,
the reconstruction error is expected to
be small~\cite{Gonstal:2025qky, Caprini:2024hue},
which makes the theoretical accuracy even more important.
At the same
time, the characteristic features of classically scale-invariant potentials
induce complications in the analysis of the phase transition from
 the phenomenological side~\cite{Ellis:2018mja,Ellis:2019oqb,Ellis:2020nnr,Athron:2022mmm},
as well as from renormalization-scale dependence and the validity of
the perturbative treatment~\cite{%
  Kierkla:2022odc,Kierkla:2023von, Kierkla:2025qyz}.

Classically scale-invariant potentials are flat around the field-space origin due to
the lack of an explicit mass term. 
This feature allows for supercooling.
The radiatively induced barrier can last until low temperatures,
delaying the onset of the phase transition. 
On the other hand, the radiatively generated minimum, whose presence is
related to the logarithmic running of the scalar coupling constant,
is separated by orders of magnitude from the barrier. 
This leads to nucleation of very thick bubbles, as the field transitions from
the metastable vacuum to a state close to
the barrier (rather than to the vicinity of the stable minimum). 
Then, even if the characteristic temperature of the phase transition
is significantly below the scale of the minimum (or the mass of the new gauge boson),
it is large compared to the field-dependent mass of the vector at
the nucleation point~\cite{Kierkla:2022odc, Kierkla:2023von}. 
Thus, for the computation of the bubble nucleation rate,
the high-temperature regime is accurate,
which leads to the necessity of thermal resummations. 
The simplest approach is the so-called daisy resummation~\cite{Parwani:1991gq,Arnold:1992rz}. 
However, to systematically improve the accuracy,
the formalism of high-temperature dimensional reduction (3D~EFT)~\cite{%
  Ginsparg:1980ef,Appelquist:1981vg,Kajantie:1995dw,Braaten:1995cm}
  can be employed. 
By including certain two-loop contributions,
belonging to the next-to-leading order (NLO) in
the so-called soft expansion, the renormalization-scale dependence of
the potential is reduced~\cite{Gould:2021oba}.

We compare~\cite{Lofgren:2023sep}
two forms of thermal resummation that are employed
for such transitions:
\begin{itemize}
\item[(i)]
  the NLO (in soft expansion) computation within the 3D~EFT constructed with
  two-loop matching~\cite{Kierkla:2023von} and functional determinants for
  the scalar and gauge sectors included in the
  exponential prefactor of the nucleation rate~\cite{Kierkla:2025qyz,Gould:2021ccf,Kierkla:2025vwp},
\item[(ii)]
  the 4D daisy-resummed effective potential.
\end{itemize}
Using both approaches, we scan the parameter space of the model spanned by
the mass and coupling of the new dark gauge boson,
$\{\gX,\MX\}$ and
determine the thermodynamic parameters of the PT and
the resulting GW spectra. 
Then, for injected signals at LISA,
following the reconstruction strategy of~\cite{Caprini:2024hue,Lewicki:2024xan},
we reconstruct the underlying
$\{\gX,\MX\}$ parameters using a Fisher-matrix analysis.
We analyze the differences in the reconstructed values,
depending on the resummation scheme used and
compare the theoretical differences to the uncertainties from reconstruction. 
We find that
the two resummation schemes are mutually incompatible,
even with careful choice of
the renormalization-group (RG) scale~\cite{Christiansen:2025xhv}.
Overall, the theoretical differences dominate over
the reconstruction uncertainty, pointing towards the necessity for
theoretical diligence, especially for strong signals.

The remainder of this \article{} is organized as follows.
Section~\ref{sec:model} introduces the conformal $\UOneX$ model and the
origin of supercooling.
Section~\ref{sec:resummation} contrasts the
two methods for computing the nucleation rate.
Section~\ref{sec:reconstruction} describes the parameter-reconstruction
procedure, and
sec.~\ref{sec:results} presents the results of the reconstruction
and compares reconstruction uncertainty
to differences between the two resummation methods.
We conclude in sec.~\ref{sec:conclusion}.
Appendix~\ref{sec:app:matching} collects the dimensional-reduction and
matching relations, and
appendix~\ref{sec:app:RG} collects the RG running.

%
\section{Conformal Abelian model and supercooling}
\label{sec:model}

\subsection{Field content and tree-level potential}
\label{sec:model:tree}

Classically conformal models do not contain a mass term in the
tree-level potential~\cite{Coleman:1973jx, Meissner:2007xv,Iso:2009ss}.
The SM Higgs mass parameter is instead generated by a portal coupling to
a new scalar $\Phi$ charged under an additional gauge symmetry.
We extend the SM by a $\UOneX$ dark sector with a gauge coupling
$\gX$~\cite{Iso:2009ss,Khoze:2014xha},
a complex scalar
$\Phi = \frac{1}{\sqrt{2}}(\phibg + \phiField  + i G)$ of charge $\QX=1$, and
an Abelian $X$ gauge boson.
Here,
$\phibg$ is the background field,
$\phiField$ the radial mode, and
$G$ the Goldstone boson.
The scalar couples to the gauge field through the covariant derivative
$D_\mu\Phi = (\partial_\mu - i\,\QX \gX X_\mu)\Phi$,
so that due to the interaction with
the background $\phibg$,
the gauge boson acquires the field-dependent mass
$\mX(\phibg) = \QX \gX \phibg$
after radiative $\UOneX$ symmetry breaking.
The scale-invariant tree-level potential reads
\begin{equation}
\label{eq:V0}
  V_\rmi{tree}(H,\Phi) =
    \lh\, (H^\dagger H)^2
  + \lphi\, (\Phi^\dagger \Phi)^2
  - \lp\, (H^\dagger H)(\Phi^\dagger \Phi)
  \,,
\end{equation}
where $H$ is the SM Higgs doublet and $\lp>0$ the portal coupling.

In the classically conformal limit, the electroweak scale and Higgs mass must be successfully reproduced via the portal term in eq.~\eqref{eq:V0}.
This leaves two free input parameters,
which we choose to be $\{\gX,\MX\}$.
For a detailed description of how to initialize the model,
see appendix~\ref{sec:app:RG}
and~\cite{Kierkla:2022odc,Kierkla:2023von}.
Typically, the barrier separating the minima is
the thinnest along the $\phi$ axis and
the transition proceeds along this direction~\cite{Prokopec:2018tnq,Kierkla:2022odc}.
Therefore, in the remaining part of the paper,
we focus on studying the phase transition in the dark sector.

\subsection{The one-loop potential and the origin of strong supercooling}
\label{sec:model:supercool}

The one-loop potential along the dark-scalar direction
receives contributions from
gauge-boson loops.
In four dimensions,
the finite part of the one-loop potential
in dimensional regularization,
using Landau gauge and
the \MSbar{} renormalization scheme,
is given by
\begin{equation}
\label{eq:V1}
  V^{(1)}(\phibg)=
    \frac{(D-1)}{2}\int_P \ln\Bigl(P^2 + \mX^2(\phibg)\Bigr)
  \stackrel{D=4-2\epsilon}{=}
    \frac{3}{64 \pi^2}\mX^4(\phibg)
    \biggl(\ln\frac{\mX^2(\phibg)}{\LamD^2}-\frac{5}{6}\biggr)
    \,,
\end{equation}
where
$D = d + 1 = 4$ and
$\int_{P} = \int \frac{\dd^D \vec{p}}{(2\pi)^D}$ and
$\mX(\phibg)=\gX\phibg$.
We neglect scalar loop contributions,
which scale as $\lphi^2\sim\gX^8$ at zero temperature
and thus contribute at $\mathcal{O}(\gX^8)$
as a higher-order effect.%
\footnote{%
  As we explain in sec.~\ref{sec:NLOdet},
  the \NLOdet{} approach includes scalar fluctuations
  through the functional determinant of the scalar sector.
} 
The thermal one-loop contribution from
the gauge field at temperature $T$ is finite and reads
\begin{align}
\label{eq:VT}
    V^{(1)}_{\T}(\phibg, T) &=
      (D-1)\, J_\T\!\left(\frac{\mX(\phibg)}{T}\right)
    \,,&
    \mathrm{with}&&
    J_\T(y) &=
      - T \int_{\vec{p}} \ln\Bigl( 1- \nB\bigl(\omega_y\bigr)\Bigr)
    \,,
\end{align}
where
$\int_{\vec{p}} = \int \frac{\dd^d \vec{p}}{(2\pi)^d}$,
$\nB(x) = \frac{1}{e^x - 1}$ is the Bose-Einstein distribution, and
$\omega_y = \sqrt{\vec{p}^2 + y^2}$.

To capture the physics across the vastly different scales present in this potential,
ranging from
the gauge-boson mass $\MX$ to
the temperature $T$,
we employ
renormalization-group~(RG) improvement
of the effective potential~\cite{Kierkla:2022odc,Kierkla:2023von}.
To this end, we set
a field-dependent RG scale taking the values
\begin{equation}
  \label{eq:mu-choice}
   \LamD = \tilde{\mu} = \max\bigl(\mX(\phibg), \LamDref\bigr)
  \,,
\end{equation}
where $\LamDref = \LamX\pi T$, with
$\LamX \in \{1/2, 1, 2\}$, is
the thermal reference scale~\eqref{eq:LamD:ref}.
At high-field values and low temperatures,
the scale is set by the field-dependent mass
$\tilde{\mu} = \mX(\phibg)$, while
at low-field values and high temperatures,
the thermal scale dominates with
$\tilde{\mu} = \LamX\pi T$.

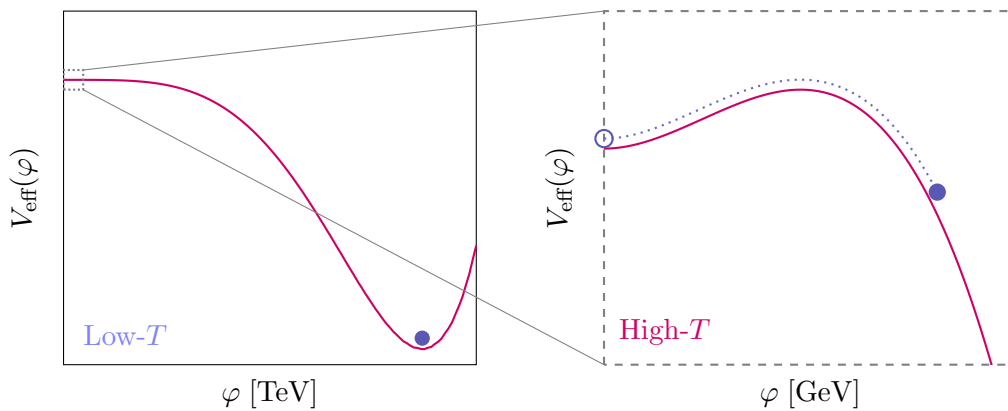
\begin{figure}[t]
  \centering
    \begin{tikzpicture}[scale=1.3]
      \draw[black] (0,0) rectangle (4.2,3.6);
      \begin{scope}
        \clip (0,0) rectangle (4.2,3.5);
        \draw[CustomRed, thick, domain=0:4.2, samples=120]
          plot (\x, {2.9 + 0.00232*\x*\x*\x*\x*\x*\x - 0.0463*\x*\x*\x*\x});
        \draw[densely dotted, thick, gray] (0, 2.8) rectangle (0.2, 3.0);
      \end{scope}
      \fill[CustomBlue!70!black] (3.65, 0.27) circle (2.2pt);
      \node[] at (2.1, -0.3) {$\phibg\;[\text{TeV}]$};
      \node[rotate=90] at (-0.4, 1.8) {$\Veff(\phibg)$};
      \node[above right] at (0.1, 0.1)
        {\color{CustomBlue} Low-$T$};
      \draw[gray, thin] (0.2, 3.0) -- (5.5, 3.6);
      \draw[gray, thin] (0.2, 2.8) -- (5.5, 0.0);
      \draw[dashed, thick, gray] (5.5,0) rectangle (9.7,3.6);
      \begin{scope}
        \clip (5.5,0) rectangle (9.7,3.6);
        \draw[CustomRed, thick, domain=5.5:9.7, samples=120]
          plot (\x, {2.2 + 0.45*(\x-5.5)*(\x-5.5) - 0.15*(\x-5.5)*(\x-5.5)*(\x-5.5)});
        \draw[CustomBlue!80!black, dotted, thick]
          plot[variable=\t, domain=0:3.3, samples=80]
          ({5.5 + \t - 0.102*(0.9*\t - 0.45*\t^2)/sqrt(1 + (0.9*\t - 0.45*\t^2)^2)},
           {2.2 + 0.45*\t^2 - 0.15*\t^3 + 0.102/sqrt(1 + (0.9*\t - 0.45*\t^2)^2)});
      \end{scope}
      \draw[CustomBlue!70!black, thick] (5.5, 2.302) circle (2.5pt);
      \fill[CustomBlue!70!black] (8.891, 1.757) circle (2.5pt);
      \node[] at (7.6, -0.3) {$\phibg\;[\text{GeV}]$};
      \node[rotate=90] at (5.05, 1.8) {$\Veff(\phibg)$};
      \node[above right] at (5.55, 0.1)
        {\color{CustomRed} High-$T$};
    \end{tikzpicture}
\caption{%
    Schematic illustration of the thermal effective potential $\Veff(\varphi,T)$
    of the conformal $\UOneX$ model along the dark scalar direction,
    illustrating the supercooled barrier and
    its scale separation from the global minimum.
}
\label{fig:potential}
\end{figure}

For $\gX\phibg \lesssim T$,
the high-temperature expansion of $J_\T$ is valid
(see e.g.~\cite{Laine:2016hma}).
In this regime, the thermal contribution grows as $T^2\gX^2\phibg^2$.
When this term is combined with the
quartic term
$\lphi(\LamD)\phibg^4$
(with $\lphi < 0$ at low enough $\LamD$),
a barrier develops that separates the symmetric
phase at $\phibg=0$ from the broken phase at $\phibg=\vphi$.
Because of the classical scale invariance of the tree-level potential,
this barrier persists down to $T=0$.
As temperature decreases, the positive thermal
term shrinks, but the barrier remains
because the quartic coupling is negative around the origin of the field space.
Consequently, the Coleman-Weinberg radiative minimum,
generated by the logarithmic running of the quartic coupling,
is separated
from the barrier by many orders of magnitude
(cf.\ fig.~\ref{fig:potential}).
This mechanism,
based on a persisting radiative barrier combined with a widely separated minimum,
is the origin of the strong supercooling characteristic of classically conformal models.

Strong supercooling has several important consequences. 
First, the nucleation temperature,
defined as the temperature at which at least
one bubble has nucleated per Hubble volume, can be
many orders of magnitude below the critical temperature.
Consequently, the phase transition occurs during a phase of
vacuum-energy-dominated thermal inflation,
where the Hubble parameter is set by
the potential difference
with the reduced Planck mass $\Mpl$,
\begin{equation}
  \label{eq:Hubble}
    H = \left(\frac{\Delta V}{3\Mpl^2}\right)^{\frac{1}{2}} \sim
    \frac{\MX^2}{\Mpl}
    \,,
\end{equation}
in contrast to the radiation-dominated case, where $H \sim T^2/\Mpl$.
The release of latent heat becomes significant, and
the transition strength is characterized by
\begin{equation}
    \label{eq:alpha}
    \alpha = \frac{\Delta V}{\rho_r}
    \,,
\end{equation}
where $\rho_r = \frac{\pi^2}{30}\geff T^4$ is
the radiation energy density,
with $\geff$ the number of relativistic degrees of freedom.
The amount of supercooling in the $\UOneX$ model is
controlled mostly by $\gX$.
The lowest values of this coupling yield the strongest,
most supercooled transitions~\cite{Ellis:2020nnr,Kierkla:2025vwp}.

To compute reliable predictions for
the thermal transition parameters,
the one-loop thermal effective potential alone is insufficient.
In the high-temperature regime, higher-loop-order diagrams become
comparable in magnitude to one-loop contributions, necessitating
a resummation~\cite{Arnold:1992rz,Parwani:1991gq}.
This is essential not only at very high temperatures.
Even during the supercooled regime,
the nucleation
occurs in regions of temperature and field space where $\gX\phibg \lesssim T$,
placing the transition in the high-temperature domain.
Furthermore, explicit calculations~\cite{Kierkla:2023von}
demonstrate that thermal resummation
significantly affects both the effective potential and
the bubble nucleation rate,
making its inclusion mandatory for precise predictions.
Section~\ref{sec:resummation} details two approaches
to implementing this resummation.

%
\section{%
  Two approaches to the computation of the bubble nucleation rate}
\label{sec:resummation}

The thermal bubble nucleation rate
\begin{equation}
\label{eq:rate}
  \Gamma(T) =
    A_\rmi{dyn}\times A_\rmi{stat} =
    A_\rmi{dyn}\times
    A_\rmi{fluc}\, e^{-\Seff}
  \,,
\end{equation}
factorizes into
a dynamical part $A_\rmi{dyn} \sim T$ and
a statistical part $A_\rmi{stat} \sim T^3$.
The dynamical part accounts for non-equilibrium evolution,
which we will not consider here.
The statistical part, which is the dominant one, 
is controlled by the effective action $\Seff$ evaluated on the classical solution
of the Euclidean equations of motion (the bounce solution) and
the fluctuation prefactor
$A_\rmi{fluc}$~\cite{%
  Langer:1969bc,Callan:1977pt,Ekstedt:2021kyx,Ekstedt:2022tqk}.

It is instructive to organize the statistical part according to
the various thermal scales that contribute to it~\cite{Gould:2021ccf}.
Denoting the energy scale relevant for nucleation by $\Lambda_\text{nucl}$,
the effective action $\Seff$ collects all contributions from the heavy modes
(e.g.\ gauge bosons) with energies $E>\Lambda_\text{nucl}$,
while the prefactor $A_\text{fluc}$ captures the light fluctuations with
$E<\Lambda_\text{nucl}$,
\begin{align}
\label{eq:Astat}
    A_\text{stat} = 
    \underbrace{
      \vphantom{A(T)}
      A_\text{fluc}
    }_{E<\Lambda_\text{nucl}} \times \,
    \underbrace{
      \vphantom{A(T)}
      e^{-\Seff}
    }_{E>\Lambda_\text{nucl}}
  \,.
\end{align}
In practice, $\Lambda_\text{nucl}$ usually coincides with the mass of
the nucleating scalar field.%
\footnote{%
  In the thin-wall regime the scale $\Lambda_\text{nucl}$
  can lie well below the scalar mass~\cite{Gould:2021ccf}.
}

At one-loop order,
$\Seff$ is the sum of the tree-level action
and the functional determinants of the fluctuations, both evaluated on
the critical-bubble background.
The prefactor $A_\rmi{fluc}$ then corresponds to the nucleating-scalar fluctuation
determinant, whose modes satisfy $E<\Lambda_\text{nucl}$,
while the remaining determinants provide one-loop corrections to $\Seff$
whenever $E>\Lambda_\text{nucl}$.
If a field other than the nucleating one becomes light,
it too should contribute to the fluctuation prefactor. 
These determinants can be evaluated numerically~\cite{%
  Gelfand:1959nq,Forman:1987gha,Kirsten:2004qv,Kirsten:2010eg} or
approximated analytically.
One such approximation is
a derivative expansion in which the determinant is written
as a series of derivative operators.
In this approach, the zeroth-order term reproduces the one-loop
contribution to the effective potential,
the Coleman-Weinberg contribution~\cite{Coleman:1973jx}. 
This is the most commonly employed approximation,
in which $\Seff$ includes only corrections to the potential.
It works well for the heavy degrees of freedom,
whose characteristic momenta are subdominant with respect
to their mass.
For light fields, the full determinants should be computed,
which explains the split in eq.~\eqref{eq:Astat}.

In this work,
we compare two approaches to the computation of the bubble nucleation rate,
which differ in the treatment of
both the effective action and the fluctuation prefactor:
\begin{itemize}

\item[{\hyperref[sec:daisy]{Approach~1}:}]
    [\Daisy{}]
    \\
    the effective action features the leading thermal resummation,
    the dynamical and fluctuation prefactors are approximated on dimensional grounds
    (cf.~sec.~\ref{sec:daisy});
    
\item[{\hyperref[sec:NLOdet]{Approach~2}:}]
    [\NLOdet{}]
    \\
    the effective action is constructed within the 
    framework of thermal EFT in a two-step procedure
    briefly reviewed in sec.~\ref{sec:NLOdet}.
    The full one-loop fluctuation determinants from
    the scalar and gauge fields are included.
\end{itemize}

\subsection{Approach 1: \Daisy{}}
\label{sec:daisy}

In the \Daisy{} approach,
the effective action is evaluated as the sum of
a canonically normalized kinetic term and the effective potential.
The latter consists of
the tree-level contribution of eq.~\eqref{eq:V0},
restricted to the dark sector,
and the one-loop vacuum and thermal contributions,
$V^{(1)}(\phibg)$ of eq.~\eqref{eq:V1} and
$V^{(1)}_{\T}(\phibg, T)$ of eq.~\eqref{eq:VT}, respectively.
Additionally,
the effective potential is supplemented by a resummation term for the heavy thermal modes,
which generate
a Debye mass for
the longitudinal vector degree of freedom~\cite{Arnold:1992rz,Parwani:1991gq}
\begin{equation}
    V^\rmii{\Daisy{}} = -\frac{1}{12\pi} ( m_{\rmii{$X$,0}}^3 -  \mX^3)
    \,,
\end{equation}
where
$\mX$ is the usual field-dependent mass,
$m_{\rmii{$X$,0}}$ is the thermally-corrected mass of the longitudinal mode and
  $m^2_{\rmii{$X$,0}}=m^2_{\rmii{$X$}}+\mD^2$,%
\footnote{%
  It is a truncation of the expression given in eq.~\eqref{eq:app:masses:U1}
}
and
$\mD$ is given by the first term of eq.~\eqref{eq:app:mD}.
In this approach, the leading IR divergences are cured and
the important effect of the thermal mass is included.
However, it is not RG-scale invariant due to
an extra scale dependence introduced by
high-temperature effects~\cite{Gould:2021oba}.

The exponential prefactor is approximated on dimensional grounds as $T^3$,
as is the dynamical part, $A_\rmi{dyn} \sim T$.
Therefore, the final expression for the bubble nucleation rate reads
\begin{equation}
\Gamma^{\rmii{\Daisy{}}}(T) =
   T^4 e^{-S_{3}^{\rmii{\Daisy{}}}/T}
  \, ,
\end{equation}
where
\begin{equation}
  \label{eq:Seff:Daisy}
S_{3}^{\rmii{\Daisy{}}} = \int_\vec{x}
  \Bigl[
      \frac{1}{2}\left(\partial_{i} \phibg\right)^2
    + V^{(1)}(\phibg)
    + V^{(1)}_{\T}(\phibg, T)
    + V^\rmii{\Daisy{}} (\phibg, T)
  \Bigr]
  \,,
\end{equation}
evaluated on the bounce solution
using $\int_\vec{x} = \int\dd^3\vec{x}$.

\subsection{Approach 2: \NLOdet{}}
\label{sec:NLOdet}

This approach closely follows the procedure
described in~\cite{Kierkla:2023von,Kierkla:2025qyz} and
applied to the $\UOneX$ model in~\cite{Kierkla:2025vwp}.

First, the nucleation EFT is defined within
the high-temperature EFT framework.
In the first step, the hard thermal Matsubara modes are resummed, and
as a result, a so-called soft 3D thermal EFT is defined~\cite{%
  Ginsparg:1980ef,Appelquist:1981vg,Farakos:1994kx,Kajantie:1995dw,
  Braaten:1995cm,Kajantie:1996mn}.
The effective theory is constructed at
two-loop level in the EFT matching~\cite{%
  Ekstedt:2022bff,Kierkla:2025vwp}.
Then, in the second step, the soft modes of the gauge fields are
integrated out at one-loop level,
as they are much heavier than the scalar field in the broken phase.
This results in the leading-order effective action of the considered EFT
\begin{align}
\label{eq:Seff:LO}
  S_{3}^\rmii{LO} = \int_{\vec{x}} \Bigl[
      \frac{1}{2} (\partial_i \phibg)^2
    + V_3^\rmii{LO}(\phibg)
    \Bigr]
    \,,
\end{align}
with
\begin{equation}
\label{eq:Veff-EFT-LO}
  V_3^{\rmii{LO}}(\phi_3) =
    \frac12 m_3^2\,\phi_3^2
  + \frac14 \lambda_3\,\phi_3^4
  - \frac{1}{12\pi}\sum_i n_i\, \bigl(m_{i,3}^2(\phi_3)\bigr)^{3/2}
  \,,
\end{equation}
where $n_i$ are the respective degrees of freedom.
The matching relations defining the 3D couplings
$m_3^2,\,\lambda_3, \dots$ are
listed in appendix~\ref{sec:app:matching:relations}.
If these matching relations are truncated at leading order,
the \Daisy{} action~\eqref{eq:Seff:Daisy} is recovered. 

The advantage of the 3D~EFT approach is that,
once the EFT is defined,
higher-order corrections can be included systematically by
computing higher-loop corrections within it.
We organize them through a strict soft expansion~\cite{%
  Hirvonen:2021zej,Lofgren:2021ogg,Ekstedt:2022zro,Gould:2023ovu,
  Ekstedt:2024etx},
expanding in the soft parameter of $\mathcal{O}(\frac{\gX}{\pi})$,
so that only some of the two-loop contributions enter at NLO.
To this end, corrections are
evaluated on the LO bounce rather than by re-solving the equations of
motion at NLO.
The spherically symmetric solution $\phi_{3,\rmii{b}}(r)$
with radial coordinate $r$, computed from the LO action via
$\nabla_r^2\phib = \partial\Veff^\rmii{LO}/\partial\phib$ with
$\phib(\infty)=0$ and $\partial_r\phib(0)=0$,
is used to evaluate the action including the NLO corrections.
Concretely, the corrections we include are
the two-loop contributions to
the effective potential $\Veff^\rmii{NLO}$~\eqref{eq:app:VeffNLO:U1} 
that scale as $\gXt^2$
(they are also needed to restore RG-invariance~\cite{Gould:2021oba}) and
a one-loop NLO correction to the kinetic
term $Z_{\phibg_3}^\rmii{NLO}$~\eqref{eq:app:Zphi3:U1} in the effective action.

The gauge bosons are \textit{scale-shifters}~\cite{Gould:2021ccf}.
They are
heavy in the bulk of the bubble, where they generate the barrier, but
light at its tail (in the symmetric phase), where they become
lighter than the nucleating scalar field.
There, the mass hierarchy on which the nucleation EFT is built
is inverted, so that the derivative expansion of their fluctuations
breaks down~\cite{Kierkla:2023von,Kierkla:2025qyz}
and the correction to the kinetic term {\em cannot} be accurately computed using
a derivative expansion.
The gauge soft modes must therefore be treated as light degrees of freedom,
with their fluctuation determinants computed exactly,
without resorting to the derivative expansion.
This leads to the final expression for the bubble nucleation rate,
adequate for high-temperature,
scale-shifting mass hierarchies~\cite{Kierkla:2025qyz}
\begin{equation}
  \label{eq:Gamma:NLOdet}
  \Gamma^{\rmii{\NLOdet{}}}(T) =
    A_{\rm dyn}\,
    \det_{\rmii{$S$}}\,
    e^{-S_{3}^{\rmii{LO}}[\phib] + C[\phib]
    -\ln\operatorname{det}_{\rmii{$V$}}\,
    -\int_{\vec{x}} V_3^{\rmii{NLO}}[\phib]}
  \, ,
\end{equation}
where
$\det_{\rmii{$S$}}$
is the one-loop scalar fluctuation determinant, and
$\det_{\rmii{$V$}}$ accounts for the vector fluctuations.
To avoid double counting,
contributions of the zero-momentum vector modes
are removed from
the LO action~\cite{Gould:2021ccf} by adding
\begin{align}
\label{eq:p0subtract}
    C[\phib] =
  - \frac{1}{12\pi} \int_{\vec{x}}
  \Bigl[
      2 \mXt^3
    + m_{\rmii{$X$,0}}^3
    - \mD^3
  \Bigr]
  \,,
\end{align}
with
$\mXt$ ($m_{\rmii{$X$,0}}$) the spatial (temporal) gauge mode
mass given in eq.~\eqref{eq:app:masses:U1}, and
$\mD$ the Debye mass of eq.~\eqref{eq:app:mD}.
In our implementation,
$\det_{\rmii{$S$}}$ is computed using
{\tt BubbleDet}~\cite{Ekstedt:2023sqc}, while the vector determinant,
which decomposes as
$\det_{\rmii{$V$}} =
\det_{\rmii{$X_0$}}\det_{\rmii{$XG$}}\det_{\rmii{$X_T$}}\det_{\rmii{$X_g$}}$
into temporal, mixed gauge-Goldstone, transverse, and ghost
contributions, is computed with an in-house code.
The exact definition of these determinants and
the details of the computation can be found
in~\cite{Ekstedt:2021kyx,Kierkla:2025qyz,Kierkla:2026kbd}.

For the dynamical prefactor $A_\rmi{dyn}$ in eq.~\eqref{eq:Gamma:NLOdet},
we use the no-damping 
approximation~\cite{Langer:1969bc,Hanggi:1990zz,Berera:2019uyp},
$A_\rmi{dyn} \simeq \sqrt{\lambda_-/(2\pi)}$ with
$\lambda_-$ the negative eigenvalue of the scalar fluctuation operator,
as implemented in {\tt BubbleDet}~\cite{Ekstedt:2023sqc}.%
\footnote{%
  This treatment neglects out-of-equilibrium and
  damping effects~\cite{Hirvonen:2024rfg,Hirvonen:2025hqn} arising
  once the thermal bath is itself driven out of equilibrium by the
  nucleation process and back-reacts on the bubble dynamics.
  Such effects have recently been addressed from first principles
  using kinetic and Wigner-function descriptions of the
  plasma~\cite{Hirvonen:2026zaq}.
}

With this procedure,
the effective action is determined to NLO accuracy,
combining the two-loop matching in the EFT with
the NLO corrections in the soft expansion,
which together maintain RG-scale invariance.
Moreover, the gauge one-loop corrections are treated without resorting to
the derivative expansion, which is important for any gauge model, where
the gauge masses vanish in the symmetric phase, and
especially important
in cases with radiative symmetry breaking where the lack of
a negative mass term for the scalar makes
the breakdown of the EFT happen sooner.
On top of that, this approach is gauge invariant~\cite{Hirvonen:2021zej,Lofgren:2021ogg}.

%
\section{Reconstructing model parameters from GW spectra}
\label{sec:reconstruction}

\subsection{From microphysics to GW spectra}
\label{sec:reconstruction:gw}

\begin{table}[t]
    \centering
    \begin{tabular}{|l|c|c|}
        \hline
         & $\gX$ & $\MX$~[GeV] \\
        \hline\hline
        Range     & $[0.55,\,0.85]$ & $[10^{4},10^{7}]$ \\
        Prior     & linear          & logarithmic        \\
        \hline
    \end{tabular}
    \caption{%
      Parameter space scan grid in
      the dark conformal Abelian Higgs model of sec.~\ref{sec:model:tree}.
      For each parameter we lay out a lattice of $\mathcal{O}(10^4)$ points
      over the indicated range.
      The parameters are input at the scale of the corresponding dark photon mass,
      $\LamDin = \MX$, and run to the thermal scale
      $\LamDref = \pi T$~\eqref{eq:LamD:ref}.
    }
    \label{tab:scan}
\end{table}
We perform a grid scan of the parameter space in the range
defined in tab.~\ref{tab:scan},
and compute the parameters of the PT
using the two methods described in
secs.~\ref{sec:daisy} and~\ref{sec:NLOdet}.
For each method, we extract
the percolation temperature $\Tstar = \Tp$, and
the inverse duration $\beta/H$
or $R H$ from the nucleation
rate~\eqref{eq:rate}~\cite{Caprini:2019egz}.%
\footnote{%
  The transition strength $\alpha$ is determined in
  the low-temperature limit as outlined in sec.~\ref{sec:model:supercool}
  and~\cite{Kierkla:2023von}.
}

The percolation temperature $\Tp$ follows from the probability
$P = \exp(-I(T))$ for a point in space to remain in
the false vacuum~\cite{Guth:1981uk,Turner:1992tz},
obtained by integrating over the bubble nucleation history
\begin{equation}
\label{eq:I(T)}
    I(T) = \frac{4\pi}{3}
    \int_T^{\Tc} \frac{{\rm d} T'}{T'^4} \frac{\Gamma(T')}{H(T')}
    \biggl(\int_T^{T'}\!{\rm d}\widetilde{T} \frac{\vw}{H(\widetilde{T})}\biggr)^3
    \,.
\end{equation}
As we consider very strong transitions with
transition strength $\alpha \gg 1$ (cf.\ eq.~\eqref{eq:alpha}),
we set
the wall velocity~\cite{%
  Cline:2021iff,Lewicki:2021pgr,Ellis:2022lft,
  Ekstedt:2024fyq,vandeVis:2025plm, Branchina:2025jou, Branchina:2025adj,
  Ekstedt:2025awx,Branchina:2026nql
  }
to
$\vw = 1$,
while the kinetic energy fraction
$K \propto \alpha_\star/(1+\alpha_\star) \simeq 1$~\cite{Steinhardt:1981ct,Espinosa:2010hh},
so that the precise value of $\alpha_\star$ becomes irrelevant.%
\footnote{%
  Henceforth, the subscript $\star$
  denotes quantities evaluated at the time of percolation.
}
The integral
$I(T)$ measures the volume of true vacuum per unit comoving volume.
We fix $\Tp$ by requiring~\cite{Enqvist:1991xw}%
\footnote{%
  Our percolation condition~\eqref{eq:percolation_condition}
  is more conservative than the popular alternative~\cite{Vinod:1971,Ellis:2018mja, Athron:2022mmm}, defined by
  $P(\Tp) = e^{-I(\Tp)} \simeq 71\%$, which corresponds to $I(\Tp) \approx 0.34$.
  This approximation is based on classical percolation theory.
  However,
  our value has a more natural interpretation when
  expansion is taken into account~\cite{Lewicki:2024sfw}.
}
\begin{equation}
  \label{eq:percolation_condition}
  I(\Tp) = 1
  \,.
\end{equation}
A subtle point is that during the vacuum-dominated phase,
the condition~\eqref{eq:percolation_condition}
alone does not guarantee that the transition completes.
To ensure this, and thereby avoid eternal
inflation~\cite{Guth:2007ng,Borde:1993xh,Guth:1980zm},
we additionally require
that the physical false-vacuum volume is shrinking at
$\Tp$~\cite{Turner:1992tz,Ellis:2019oqb},
\begin{equation}
\label{eq:condition_false_vacuum_decreasing}
    \frac{1}{V_\mathrm{false}} \frac{\mathrm{d} V_\mathrm{false}}{\mathrm{d}t} =
    H(T) \left(3 + T \frac{\mathrm{d}I(T)}{\mathrm{d}T}\right)\biggr|_{T=\Tp} < 0
    \,.
\end{equation}
If this condition is violated, the universe keeps inflating and
the transition cannot complete at the estimated $\Tp$.

The mean bubble separation at percolation is then obtained via
the cubic root of
the inverse of the bubble number density~\cite{Turner:1992tz,Enqvist:1991xw}
\begin{equation}
\label{eq:R_star}
    R = 
    n_\rmii{$B$}^{-1/3} =
    \biggl[\Tp^3 \int_{\Tp}^{\Tc}
      \frac{{\rm d}T'}{T'^4} \frac{\Gamma(T')}{H(T')} e^{-I(T')}\biggr]^{-{\frac{1}{3}}}
      \, .
\end{equation}

Given the thermodynamic parameters,
we can determine the geometric parameters of
GW spectra and use existing templates~\cite{Lewicki:2022pdb, Lewicki:2025hxg}.
For the strongly supercooled transitions considered here,
with $\alpha \gg 1$,
the vacuum energy released during the transition is shared between
the scalar-field bubble walls and the surrounding fluid.
The GW signal is therefore sourced by
bubble-wall collisions and the motion of relativistic fluid shells,
which produce the same spectral shape~\cite{Lewicki:2022pdb}.

Following~\cite{Caprini:2024hue},
the present-day signal close to the peak is described by
a broken power law
\begin{align}
\label{eq:OmegaGW}
  \Omega_\rmii{GW}(f)\, h^2 &=
    \Omega_\rmi{p}\, S(f/\fp)
  \,,&
  S(x) =
    \frac{(a+b)^c}{\bigl[\, b\, x^{-a/c} + a\, x^{b/c} \,\bigr]^c}
  \,,
\end{align}
where the spectral shape is normalized to $S(1)=1$, with
the low- and high-frequency slopes fixed by $a,b>0$ and the width of
the peak by $c>0$.
The peak amplitude and frequency today are~\cite{Lewicki:2025hxg}
\begin{align}
\label{eq:Omega:freq:peak}
    \Omega_\rmi{p} &= \frac{1.6\times 10^{-5}}{h^2}
    \biggl[\frac{\geff}{100}\biggr]
    \biggl[\frac{\heff}{100}\biggr]^{-\frac43} A
      \,,\\[2mm]
  f_{{\rm p}} &=
    2.6\!\times\! 10^{-8}\,{\rm Hz}\,
    \frac{\Treh}{{\rm GeV}}
    \biggl[\frac{\geff}{100}\biggr]^{\frac12}
    \biggl[\frac{\heff}{100}\biggr]^{-\frac13}
    \frac{2\pi f_{{\rm p},\star}}{a_\star \Hstar}
  \,.
\end{align}
The relativistic degrees of freedom at the reheating temperature $\Treh$
are $\geff$ and $\heff$ for the energy and entropy densities, respectively.
The amplitude and frequency further depend
on $\beta$ as
\begin{align}
    A(\beta) & \approx 0.06
      \biggl[1 + 0.8 \Bigl(1-e^{\sqrt{\Hstar/\beta}} \Bigr) \biggr]
    \,,&
    \frac{2\pi f_{{\rm p},\star}(\beta)}{a_\star\Hstar} & \approx
      0.7 \frac{\beta}{\Hstar}
      \biggl[ 1 + 1.8 \Bigl(\frac{\beta}{\Hstar}\Bigr)^{-1.2} \biggr]
    \,,
\end{align}
and the dimensionless Hubble constant is
$h \simeq 0.6737 \pm 0.0054$~\cite{Planck:2018vyg}.
To compute the transition timescale,
we use the simple relation~\cite{Caprini:2019egz}
\begin{align}
\label{eq:RvsBeta}
  R &\simeq \frac{(8\pi)^{\frac{1}{3}}}{\beta}
  \,.
\end{align}
The spectral shape coefficients $\{a,b,c\}$
of eq.~\eqref{eq:OmegaGW} are also taken
from~\cite{Lewicki:2025hxg}; for the dissipative bulk-flow model, they read
$a = b = c = 2$.
To fully reproduce the spectral shape from~\cite{Lewicki:2025hxg},
we should also include the shift to
the $f^3$-tail at low frequencies corresponding to super-horizon scales.
However, this would unnecessarily complicate our analysis,
which does not attempt to reconstruct the spectral shape.
Signals of interest to us lie within the sensitivity band of LISA, and
we focus on reconstruction of the peak amplitude and frequency.

Thus, we also neglect the impact of the potential matter-domination periods
induced by the oscillating scalar after percolation.
Such a phase would generate additional features in the GW signal associated with
the decay rate $\phiField \to \mathrm{SM}$~\cite{Ellis:2020nnr}.
This would allow for a reconstruction of
the portal coupling $\lp$
in eq.~\eqref{eq:V0}~\cite{Gonstal:2025qky},
but would leave our conclusions unchanged otherwise.

Using the grid scan of the parameter space $\{\gX,\MX\}$
in tab.~\ref{tab:scan}, and following the procedures described
above, we construct a map
$\{\gX,\MX\}\mapsto\{\Omega_\rmi{p},\fp\}$.
This map is then used to reconstruct the fundamental parameters
from a given spectrum, following the algorithm described in the
next section.

\subsection{From GW spectra back to microphysics}
\label{sec:reconstruction:fisher}

In line with
the strategy of~\cite{Caprini:2024hue,Lewicki:2024xan},
we use the GW spectrum to reconstruct the fundamental input parameters
$\{\gX,\MX\}$ of the conformal $\UOneX$ model.

The Fisher matrix is given by
\begin{align}
\label{eq:fisher}
  \Gamma_{ij} &=
    \mathcal{T}\!\int\!
    \frac{\dd f}{\Omega_\rmi{tot}(f)^2}
    \frac{\partial \Omega_\rmii{GW}}{\partial\theta_i}
    \frac{\partial \Omega_\rmii{GW}}{\partial\theta_j}
  \,,
\end{align}
where we use the mapping of the parameters described above to
treat the signal as a function of fundamental model parameters $\theta_i \in \{\gX,\MX\}\,$.
The mission time is
$\mathcal{T}=4\,\rm{yr}$ and
the total signal%
\footnote{%
  For simplicity,
  we neglect the astrophysical foregrounds in our analysis.
  Their impact and a much more detailed description of
  the uncertainty estimates can be found in
  sec.~\ref{sec:intro} and in
  a dedicated LISA study~\cite{Caprini:2024hue}.
}
$\Omega_\rmi{tot}(f) = \Omega_\rmii{GW}(f) + \Omega_\rmi{instr}(f)$,
where $\Omega_\rmi{instr}$ is the instrumental noise~\cite{LISACosmologyWorkingGroup:2022jok}.

The covariance matrix $\sigma$ is the inverse of the Fisher matrix,
\begin{align}
  \sigma_{ij}^2 = (\Gamma^{-1})_{ij}
  \,,
\end{align}
from which the marginalized errors on individual parameters are extracted as
\begin{align}
  \label{eq:sigma:params}
  \sigma_{\gX} &= \sqrt{\sigma^2_{\gX\gX}}
  \,,&
  \sigma_{\MX} &= \sqrt{\sigma^2_{\MX\MX}}
  \,,
\end{align}
where $\sigma_{\gX\gX}$ and $\sigma_{\MX\MX}$ are the diagonal elements of the covariance matrix
corresponding to the gauge coupling and mass, respectively.

%
\section{Error budget of parameter space reconstruction}
\label{sec:results}

To determine the largest source
of uncertainty in the reconstruction of the underlying model parameters
from an injected LISA signal,
we compare the resummation methods from
secs.~\ref{sec:NLOdet}
and~\ref{sec:daisy}.
By focusing on the conformal $\UOneX$ model,
we conduct extensive parameter scans with
$\mathcal{O}(10^4)$ $\{\gX,\MX\}$ pairs and
three different reference four-dimensional RG scales%
\footnote{
  In the \Daisy{} approach, this scale is the reference scale
  in eq.~\eqref{eq:mu-choice},
  whereas in the \NLOdet{} approach it is the scale
  at which the matching is performed.
}
\begin{align}
  \label{eq:LamD:ref}
  \LamDref &= \LamX\pi T
  \,, &
  \LamX &\in\{1/2,1,2\}
  \,,&
  \Lamd &= T
  \,,
\end{align}
where
$\Lamd$ is the three-dimensional RG scale
present in the matching relations and
the NLO effective potential of appendix~\ref{sec:app:matching}.
Its value is fixed as in e.g.~\cite{Niemi:2021qvp,Lewicki:2024xan,Niemi:2024vzw}.
In sec.~\ref{sec:results:scaledep},
we then inject
two representative signals into LISA and reconstruct the underlying model parameters.

During the reconstruction of
the dark gauge coupling $\gX$,
we define the following
three uncertainties
\begin{align}
  \text{method uncertainty:}&& 
  \delta_g^\rmi{meth} &\equiv
    \frac{\bigl|\gX^\rmii{\Daisy{}} - \gX^\rmii{\NLOdet{}}\bigr|}{\gX^\rmii{\NLOdet{}}}
  \,,\\
  \text{reconstruction uncertainty:}&& 
  \delta_g^\rmi{exp} &\equiv \frac{\sigma_{\gX}}{\gX}
  \,,\\
  \text{renormalization scale uncertainty:}&& 
  \delta_g^{\LamD,\rmi{\tt i}} &\equiv
    \frac{\Delta_{\gX}^{\LamD,\rmi{\tt i}}}{\gX}
  \,,
\end{align}
where
$\gX^\rmii{\Daisy{}}$ and
$\gX^\rmii{\NLOdet{}}$
denote the central values of the coupling reconstructed from
the GW spectrum
using the parameter-space scan of
tab.~\ref{tab:scan} with either
the \Daisy{} or \NLOdet{} method, respectively.
Here,
$\sigma_{\gX}$ is the Fisher-matrix uncertainty on the gauge coupling,
defined in sec.~\ref{sec:reconstruction:fisher}, and
$\Delta_{\gX}^{\LamD,\rmi{\tt i}} = |\gX^\rmi{\tt i}(\LamDref^\rmi{min}) - \gX^\rmi{\tt i}(\LamDref^\rmi{max})|$
with ${\tt i} \in \{\Daisy{},\NLOdet{}\}$
quantifies the variation of the reconstructed gauge coupling
under changes in the four-dimensional reference RG scale $\LamDref$
as in eq.~\eqref{eq:LamD:ref}.
In the following, we discuss these uncertainties
and identify
a hierarchy between them in sec.~\ref{sec:results:implications}.

\subsection[Transition observables across the parameter space]{%
  Transition observables across the parameter space%
}

%
\begin{figure}[t]
\centering
\includegraphics[width=0.5\textwidth]{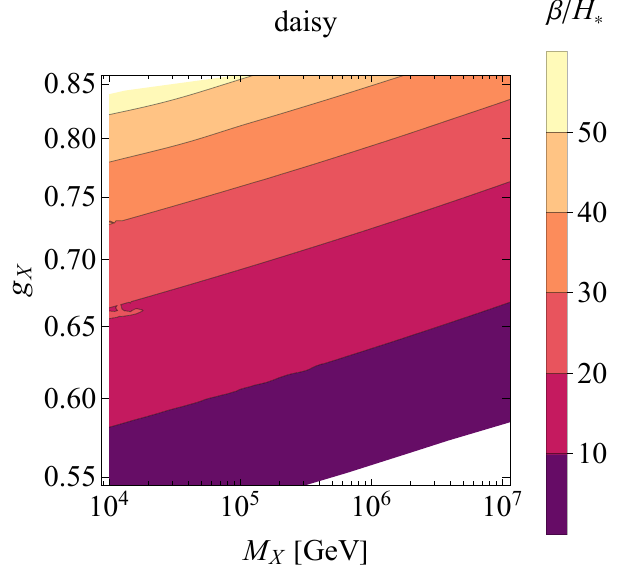}%
\includegraphics[width=0.5\textwidth]{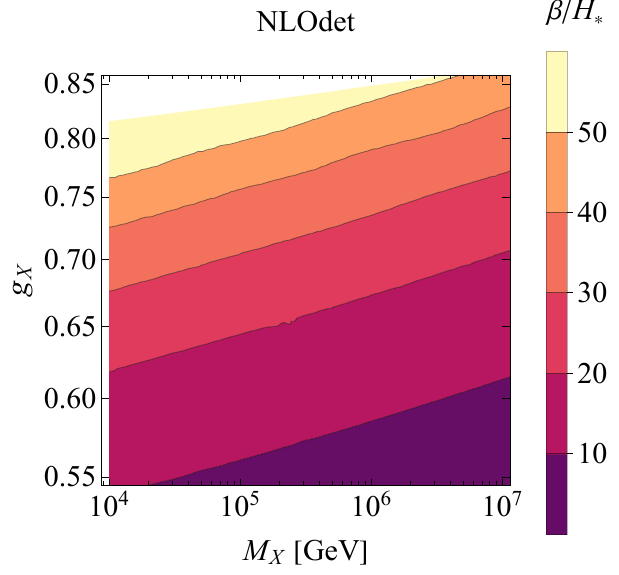}
\caption{%
  Transition inverse timescale $\beta/\Hstar$ computed from
  the mean bubble separation $R\Hstar$
  using eq.~\eqref{eq:RvsBeta}
  of the conformal $\UOneX$ transition
  over the $\{\gX,\MX\}$ parameter space, for the
  \Daisy{} (left, cf.\ sec.~\ref{sec:daisy}) and
  \NLOdet{} (right, cf.\ sec.~\ref{sec:NLOdet})
  methods.
  The lower limit on the parameter space is set by
  $\Tp < T_\rmii{QCD}$.
  }
\label{fig:RH}
\end{figure}
\begin{figure}[t]
\centering
\includegraphics[width=0.5\textwidth]{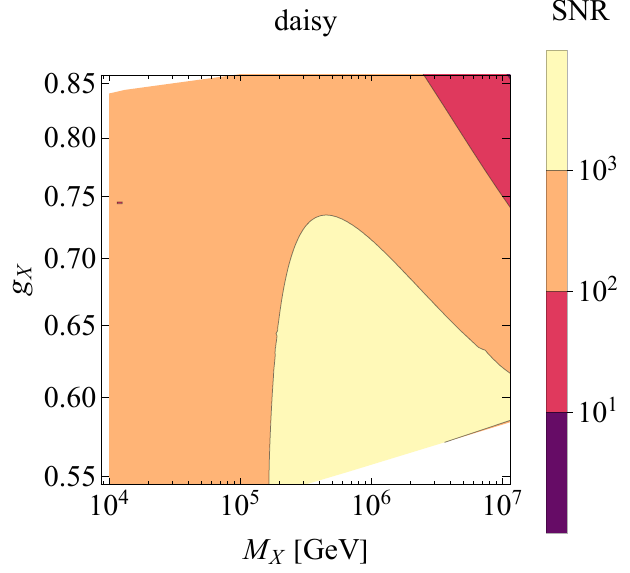}%
\includegraphics[width=0.5\textwidth]{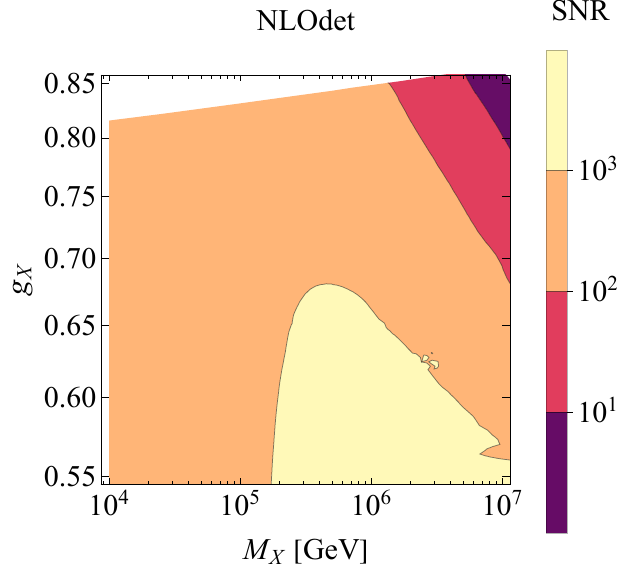}
\caption{%
  LISA signal-to-noise ratio~(SNR) of
  the conformal~$\UOneX$ transition over
  the $\{\gX,\MX\}$ parameter space, for
  \Daisy{} (left, cf.\ sec.~\ref{sec:daisy}) and
  \NLOdet{} (right, cf.\ sec.~\ref{sec:NLOdet})
  methods.
}
\label{fig:snr}
\end{figure}
First, we give an overview of the scanned parameter space at
a fixed RG scale.
In figs.~\ref{fig:RH} and~\ref{fig:snr},
we show the results of the $\LamDref = \pi T$ scan
over the parameter range of tab.~\ref{tab:scan}.
This range corresponds to the strongest transitions with
$\alpha \gtrsim \mathcal{O}(1)$, which bounds the parameter space from above,
whereas the reconstruction of weaker transitions
has been investigated in~\cite{Lewicki:2024xan}.
The lower limit is set by requiring $\Tp > T_\rmii{QCD}$ and that
the false vacuum shrinks at $\Tp$~(see~eq.~\eqref{eq:condition_false_vacuum_decreasing}).
For smaller gauge couplings, the cosmic QCD transition occurs before bubble percolation, and quark condensation accelerates
the conformal phase transition~\cite{%
  vonHarling:2017yew,Iso:2017uuu,Sagunski:2023ynd,Schmitt:2024pby}.
This requires the inclusion of QCD-induced effects in the effective potential,
which is beyond the scope of this work.
Regarding $\MX$, we set the lower limit to remain consistent with
existing collider limits~\cite{ALEPH:2006tnd,Robens:2015gla,Ilnicka:2018def,Gouttenoire:2023pxh}, while
the upper limit is chosen so that the relevant frequency regime is within the LISA band.

In the parameter space considered,
both methods yield inverse transition timescales in the range
$5 \lesssim \beta/\Hstar \lesssim 60$~(cf.~fig.~\ref{fig:RH}).
While both methods show the same qualitative behavior, the results differ quantitatively.
Phase transitions with percolation above
the QCD phase transition are predicted for $\gX \gtrsim 0.55$ using \Daisy{} resummation,
while for
the \NLOdet{} they appear down to $\gX \gtrsim 0.5$.
This corresponds to a shift of
the boundary of the allowed parameter space of
$\mathcal{O}(10\%)$ in the gauge coupling. 
Interestingly, this shift is an order of magnitude larger than the percent-level shift found in
the real scalar gauge singlet-extended SM~\cite{Lewicki:2024xan}.
This underlines the importance of computational diligence in radiatively induced FOPTs.
Note that switching between resummation methods does not shift
the range of $\MX$ significantly.
Changing $\MX$ mostly rescales the Hubble rate
during thermal inflation~\eqref{eq:Hubble},
which is evaluated in the low-temperature limit
(cf.\ fig.~\ref{fig:potential}) and is therefore
insensitive to the precise form of the nucleation rate,
so that both methods yield similar $\MX$.

Both methods predict very high SNRs at LISA,
computed following~\cite{Schmitz:2026sur} (cf.~fig.~\ref{fig:snr}),
indicating excellent observational prospects across the entire parameter space.
As for $\beta/\Hstar$, the two methods again differ by an
$\mathcal{O}(10\%)$ shift in the gauge coupling.
We next quantify how
this shift and
the residual RG-scale dependence of each method propagate into
the reconstruction of the underlying model parameters.

\subsection{Reconstructing model parameters}
\label{sec:results:scaledep}

One virtue of the \NLOdet{} approach is that the RG-scale dependence
cancels between the various contributions, to the order of the
computation~\cite{Gould:2021oba, Kierkla:2023von}.
The \Daisy{}
approach, by contrast, misses the two-loop contributions needed to
cancel the RG-scale dependence arising from high-temperature
effects~\cite{Gould:2021oba}.

\begin{figure}[t]
\centering
\includegraphics[width=0.42\linewidth]{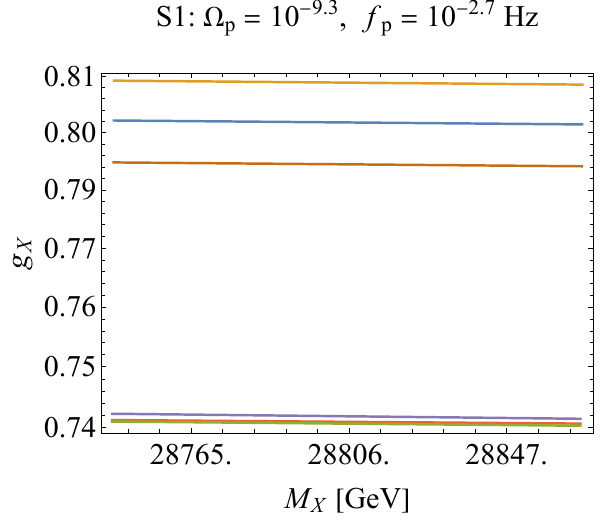}%
\includegraphics[width=0.42\linewidth]{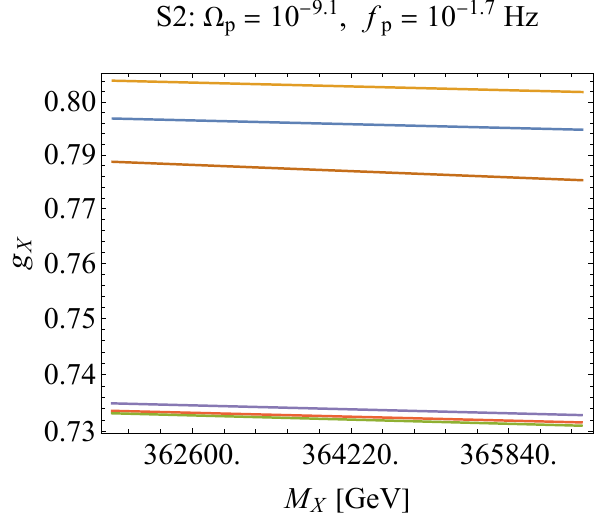}%
\hfill
\raisebox{1.3cm}{%
\includegraphics[width=0.15\linewidth]
  {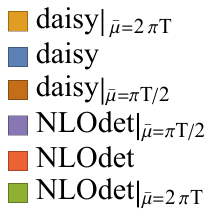}
}
\caption{%
  Fisher reconstruction ellipses
  at $2\sigma$ confidence level in the $\{\gX,\MX\}$ plane for the
  two injected signals
  {\tt S1} (left) and
  {\tt S2} (right)
  comparing the
  \NLOdet{} and \Daisy{} methods
  for different choices of the RG scale $\LamDref = \LamX \pi T$,
  $\LamX \in \{1/2,1,2\}$.
  }
\label{fig:ellipses}
\end{figure}%
We quantify the RG-scale dependence of both methods by varying the
reference scale $\LamDref$ over the canonical factor-of-two
range in eq.~\eqref{eq:LamD:ref}
and tracking the induced shift in $\{\Tstar,\beta/\Hstar\}$%
\footnote{%
  We do not track $\alpha$ here, as it is evaluated at
  $\MX$ and is therefore, by construction, insensitive to
  variations of $\LamDref$.
}
and, ultimately,
in the reconstructed $\{\gX,\MX\}$,
shown in fig.~\ref{fig:ellipses}.
The panels correspond to
the two injected LISA signal examples
{\tt S1} (left) and
{\tt S2} (right).
These points are chosen as representative targets with strong
signals in the LISA band at two distinct peak frequencies, allowing
us to test the robustness of the reconstruction across the
frequency range relevant for LISA.

\begin{figure}[t]
\centering
\includegraphics[width=0.5\textwidth]{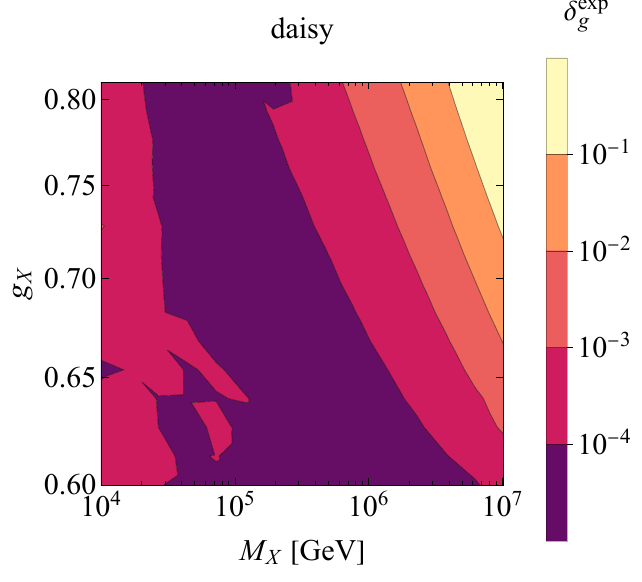}%
\includegraphics[width=0.5\textwidth]{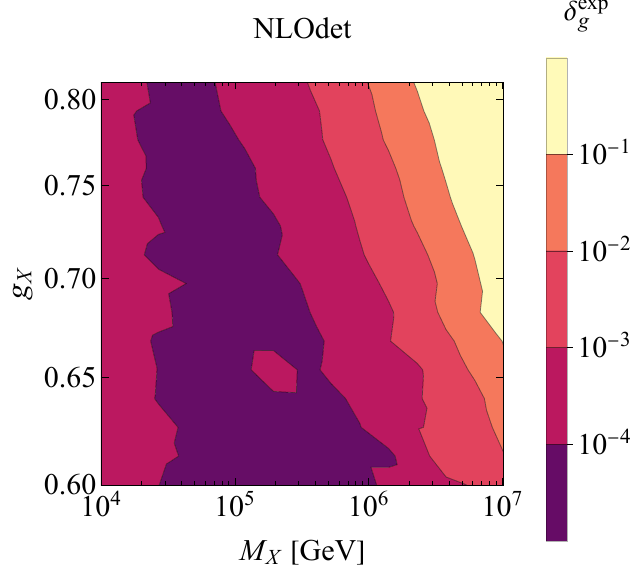}
\caption{%
  Relative Fisher uncertainty
  $\delta_{g}^{\rmi{exp}} = \sigma_{\gX}/\gX$ on
  the reconstructed gauge coupling over the $\{\gX,\MX\}$ parameter space,
  for the
  \NLOdet{} (left) and \Daisy{} (right) methods.
  }
\label{fig:sigma_gg}
\end{figure}
As a consequence of the high amplitude of the GW signal,
the parameter reconstruction is extremely accurate throughout the scan.
The resulting reconstructed
$2\sigma$ confidence ellipses are correspondingly narrow,
with sub-percent uncertainty bands,
$\delta_g^\rmi{exp} < \mathcal{O}(1\%)$,
in both reconstructed parameters $\gX$ and $\MX$.

The residual scale dependence for \NLOdet{} is barely
discernible at the plotting level.
For \Daisy{},
we find that changing the RG scale by
a factor of two shifts the reconstructed gauge coupling by
$\Delta_{\gX}^{\LamD,\rmii{\Daisy{}}} \sim \mathcal{O}(0.01)$,
corresponding to a
$\delta_g^{\LamD,\rmii{\Daisy{}}} = \mathcal{O}(1\%)$ uncertainty.
This shift, however, is an order of magnitude larger than
the confidence intervals of the reconstruction,
leading to clearly separated ellipses corresponding to different RG scales. 
Therefore, the theoretical error from incomplete resummation 
clearly dominates over the experimental uncertainties,
$\delta_g^{\LamD,\rmii{\Daisy{}}} \gg \delta_g^\rmi{exp}$.

Most importantly,
the gauge couplings reconstructed using \Daisy{} deviate by
$\delta_g^\rmi{meth} = \mathcal{O}(10\%)$ from the results obtained via \NLOdet{}. 
This is an order of magnitude larger than the scale dependence of \Daisy{}.
None of the reasonable choices of
$\LamDref$ of eq.~\eqref{eq:LamD:ref}
brings the reconstructed value of 
$\gX^\rmii{\Daisy{}}$ within the $1\sigma$ ellipse of $\gX^\rmii{\NLOdet{}}$. 
While RG-scale dependence is typically used as
an indicator of the impact of missing higher-order corrections,
we show that, in conformal models,
RG-scale variation does not adequately capture the effects of
a full NLO analysis including fluctuation determinants.
This is in contrast to a recent work~\cite{Christiansen:2025xhv} that argued that
a shift of the RG scale
brings the \Daisy{} prediction into agreement with the higher-order results. 

Performing the reconstruction for points across the parameter space
as in fig.~\ref{fig:ellipses},
with the injected spectra at each point set by the map $\{\gX,\MX\}\mapsto\{\Omega_\rmi{p},\fp\}$,
we obtain the relative Fisher uncertainty
$\delta_g^\rmi{exp} = \sigma_{\gX}/\gX$ for both methods.
The results are shown in fig.~\ref{fig:sigma_gg}.
The Fisher uncertainty remains at
the sub-percent level across most of the parameter space and closely tracks the SNR of fig.~\ref{fig:snr},
being smallest where the signal is strongest and
growing towards the weaker-signal region.
Both methods yield similar uncertainty maps,
since $\delta_g^\rmi{exp}$ is set by the signal strength and the detector sensitivity.
The only difference between the methods is therefore
the shift in $\gX$ observed earlier.

\subsection{Uncertainty hierarchy of parameter reconstruction}
\label{sec:results:implications}

\begin{figure}[t]
\centering
\includegraphics[width=0.7\textwidth]{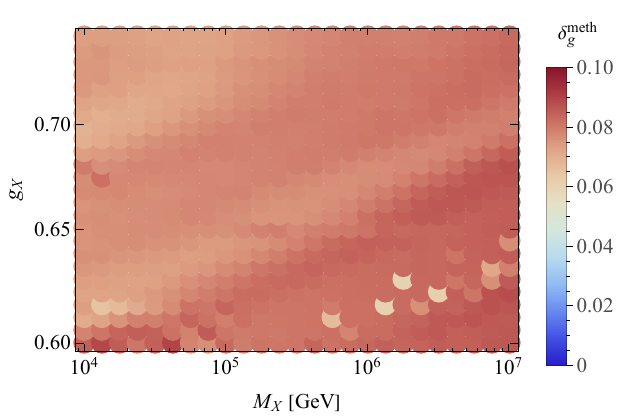}
\caption{%
  Relative methodological difference
  $\delta_g^\rmi{meth} \sim \mathcal{O}(10\%)$ in
  the reconstructed gauge coupling $\gX$ between
  the \Daisy{} and \NLOdet{} methods at the fixed central
  RG scale $\LamX=1$ of eq.~\eqref{eq:LamD:ref},
  evaluated point by point across the parameter space.
  The scattered points at small $\gX$ and large $\MX$
  reflect local numerical instabilities in computing the underlying thermodynamic parameters.
  }
\label{fig:delta_gg}
\end{figure}
We compute the relative methodological difference $\delta_g^\rmi{meth}$ by
finding the shift in $\gX$ required for the two methods to produce the same GW spectrum.
As is evident from fig.~\ref{fig:ellipses},
no accompanying shift in $\MX$ is required,
which we verified across the rest of the parameter space.
The resulting $\delta_g^\rmi{meth}$ is shown over the $\{\gX,\MX\}$ parameter space
in fig.~\ref{fig:delta_gg}.

The discrepancy stays consistently at the
$\delta_g^\rmi{meth} = \mathcal{O}(10\%)$ level,
with only a mild dependence on $\gX$ and $\MX$,
confirming that the shift observed for
the two injected signals
{\tt S1} and {\tt S2} in fig.~\ref{fig:ellipses} is generic
and not specific to those points.
Comparing with the sub-percent Fisher uncertainty of fig.~\ref{fig:sigma_gg},
the method difference exceeds the experimental one by
more than an order of magnitude
whenever the signal is within the LISA band
and therefore lies well above both the scale and reconstruction uncertainties.

Our results establish a clear hierarchy between experimental, scale, and methodological uncertainties:
\begin{equation}
\label{eq:uncertainty:hierarchy}
  \delta_g^\rmi{exp}
  \sim
  \delta_g^{\LamD,\rmii{\NLOdet{}}}
  \ll
  \delta_g^{\LamD,\rmii{\Daisy{}}}
  \ll
  \delta_g^\rmi{meth}
  \sim \mathcal{O}(10\%)
  \,.
\end{equation}
From the uncertainty hierarchy~\eqref{eq:uncertainty:hierarchy},
it becomes evident that
for strongly supercooled conformal transitions,
the dominant limitation of parameter reconstruction is theoretical.
The choice of
the method used to compute the bubble nucleation rate
dominates over the detector sensitivity.
This places precise calculations 
including consistent resummations and fluctuation determinants at
the center of any precision GW inverse program in these models,
on top of the reconstruction-oriented program of~\cite{Caprini:2024hue,Lewicki:2024xan}.

These results apply to the LISA band, where
tab.~\ref{tab:scan} yields $\fp \gtrsim 10^{-4}$~Hz.
The nHz PTA band would instead require much lighter gauge bosons,
$\MX \sim \mathcal{O}(1)$~GeV~\cite{Christiansen:2025xhv},
which in our model is only viable in a completely decoupled conformal
sector, as it would otherwise
be strongly constrained through collider searches~\cite{ALEPH:2006tnd,Robens:2015gla,Ilnicka:2018def,Gouttenoire:2023pxh}.
There, the methodological difference should persist at the
$\delta_g^\rmi{meth} \sim \mathcal{O}(10\%)$ level,
while the reconstruction uncertainty grows substantially due to the
lower SNR, so that the two uncertainties become comparable,
$\delta_g^\rmi{exp} \sim \delta_g^\rmi{meth}$, and
the hierarchy~\eqref{eq:uncertainty:hierarchy}
does not hold for the currently available data~\cite{NANOGrav:2023gor,EPTA:2023fyk,Reardon:2023gzh,Xu:2023wog}.

%
\section{Conclusions}
\label{sec:conclusion}

In the context of strongly supercooled first-order phase transitions
in the classically conformal $\UOneX$ extension of the SM,
we determined
the uncertainty of
reconstructing model parameters from
an injected GW spectrum at LISA.
We then compared
this uncertainty to the error induced by
an incomplete treatment of the bubble nucleation rate.
To this end, we contrasted two methods:
the
\NLOdet{}
method, which incorporates both
the resummation of heavy thermal modes at NLO and
the full fluctuation determinants in the nucleation rate,
and
the \Daisy{} method, which uses
the 4D daisy-resummed effective potential.

The \NLOdet{} approach 
respects the thermal scale hierarchy present in
the high-temperature regime, and is thus applicable to supercooled
potentials around the barrier.
It mitigates the RG-scale dependence through the inclusion of
NLO contributions and, via the full fluctuation determinants,
accounts for the scale-shifting nature of the gauge fields.
This enables a consistent inclusion of NLO corrections to
the kinetic term in the effective action, together with the scalar
fluctuations around the bounce.
Finally, the approach is gauge invariant.
The \Daisy{} approach, by contrast,
is the simplest way of resumming some
high-temperature fluctuations, but it satisfies none of
the above
properties~\cite{Lofgren:2023sep,Kierkla:2025qyz,Ekstedt:2021kyx}.

Reconstructing the fundamental parameters $\{\gX,\MX\}$ from
an injected LISA signal,
we found that
\begin{itemize}
    \item[$\lvert\,\mathbf{1}\,\rvert$]
        the two analyzed schemes predict
        distinct regions of parameter space that
        {\em cannot} be reconciled by a simple shift of
        the RG scale; see e.g.~\cite{Christiansen:2025xhv}.
        The residual discrepancy is
        rooted in the absence of physical higher-order contributions
        rather than in an improper choice of RG scale;
    \item[$\lvert\,\mathbf{2}\,\rvert$]
        quantitatively, the two schemes disagree by
        $\mathcal{O}(10\%)$ in the reconstructed gauge coupling $\gX$,
        whereas the residual RG-scale dependence amounts to an
        $\mathcal{O}(1\%)$ shift for \Daisy{} and
        is negligible for \NLOdet{};
    \item[$\lvert\,\mathbf{3}\,\rvert$]
        the theoretical error of the \Daisy{} method
        thus exceeds
        the sub-percent experimental error of the Fisher reconstruction, so
        that the accuracy of
        the parameter reconstruction is limited by
        the theoretical framework rather than
        by the detector sensitivity, as summarized by the
        hierarchy~\eqref{eq:uncertainty:hierarchy}.
\end{itemize}
The reconstructed mass $\MX$, by contrast, is essentially insensitive to
the treatment of the nucleation rate,
as it mostly sets the Hubble scale at low temperatures.
This hierarchy is specific to the high-SNR LISA band.
In the PTA band, reached only for much lighter gauge bosons
$\MX \sim \mathcal{O}(1)$~GeV,
the methodological uncertainty is expected to
become comparable to that from the reconstruction procedure,
given the still significant experimental uncertainties
of PTA data.

Our findings apply to
the so-called GW {\em inverse problem}~\cite{%
  Hashino:2016xoj,Friedrich:2022cak,Caprini:2024hue,Lewicki:2024xan},
more broadly.
Whenever a strong, well-measured GW signal is used to reconstruct the
underlying particle-physics model, the theoretical uncertainty of the
nucleation-rate calculation must be controlled at least as tightly as
the experimental one, since otherwise it dominates the error budget.
For the strongly supercooled transitions of classically conformal
models, this favors the \NLOdet{} approach,
which resolves many of the theoretical issues listed above.
Scans of the non-Abelian $\SUTwoX$ extension
of~\cite{Kierkla:2025qyz,Kierkla:2025vwp} display the
same qualitative pattern, suggesting that the effect is generic to
radiatively induced transitions.
A natural next step is to assess the impact of higher-dimensional
operators in the nucleation EFT, which we leave for future work.

%
\section*{Acknowledgements}

We thank
Aleksandr Azatov,
Andreas Ekstedt,
Oliver Gould,
Joonas Hirvonen,
Cristina Puchades-Ib\'a\~nez,
Miha Nemev\v{s}ek,
Nicklas Ramberg,
Tuomas V.I. Tenkanen, and
Jorinde van de Vis
for illuminating discussions.

MK is supported by the Carl Trygger Foundation through
grant no.~{\tt CTS~24:3412}.
ML is supported by the National Science Centre, Poland,
grant no.~{\tt 2023/50/E/ST2/00177},
and by the FNP IRA programmes: AstroCeNT (MAB/2018/7), funded from the ERDF; and
Astrocent Plus (FENG.02.01-IP.05-A015/25), co-financed by
the European Union under
FENG 2021--2027; and
the Teaming for Excellence grant Astrocent Plus (101137080),
funded by the European Union with
complementary national funding from the MNiSW
(MNiSW/2025/DIR/811).
PS is supported by the Swiss National Science Foundation (SNSF) under grant
\href{https://data.snf.ch/grants/grant/215997}{\tt PZ00P2-215997}.
DS is supported by the DFG through the Emmy Noether Programme, project
no.~\href{https://gepris.dfg.de/project/548044346/2?lang=en}{\tt 548044346}.
B\'S is supported by the National Science Centre, Poland, through the OPUS grant
no.~{\tt 2023/49/B/ST2/02782}.
\\

\noindent
{\bf Data availability statement.}
For the purpose of Open Access, the authors have applied
a CC-BY public copyright licence to any Author Accepted Manuscript
version arising from this submission.
The data used to prepare
figs.~\ref{fig:RH}--\ref{fig:delta_gg} presented in this article are
publicly available at~\cite{ZenodoData}.
The effective potential expressions
are publicly available via the software
{\tt DRalgo}~\cite{Ekstedt:2022bff}.

%
\appendix
\renewcommand{\thesection}{\Alph{section}}
\renewcommand{\thesubsection}{\Alph{section}.\arabic{subsection}}
\renewcommand{\theequation}{\Alph{section}.\arabic{equation}}

%
\section{Dimensional reduction and matching relations}
\label{sec:app:matching}

This appendix collects the high-temperature dimensional-reduction setup and the matching relations of the effective parameters entering
the nucleation rate of the conformal $\UOneX$ model.
The relations below follow the construction
of~\cite{Kierkla:2023von,Schicho:2021gca} and can be reproduced with
{\tt DRalgo}~\cite{Ekstedt:2022bff}.

\subsection{Matching relations}
\label{sec:app:matching:relations}

The matching relations below extend those
of~\cite{Hirvonen:2021zej} to the conformal $\UOneX$ model and were
obtained both with in-house {\tt FORM}~\cite{Davies:2026cci} software and
with {\tt DRalgo}~\cite{Ekstedt:2022bff}.
To one-loop order in the high-temperature expansion, the soft 3D gauge
coupling and Debye mass read
\begin{align}
\label{eq:app:gX3}
  \gXt^{2} &=
    \gX^2 T\Bigl[
      1 - \frac{\gX^2}{(4\pi)^2}\frac{\Lb}{3}
    \Bigr]
  \,,\\
\label{eq:app:mD}
  \mD^{2} &=
    \frac{\gX^{2} T^{2}}{3}\Bigl[
      1 - \frac{1}{(4\pi)^2}\Bigl(
          \frac{\Lb - 7}{3}\,\gX^{2}
        - 4 \lambda_{\phi}
      \Bigr)
    \Bigr]
  \,,
\end{align}
with the shorthands
\begin{align}
\label{eq:app:Lb}
  \Lb &\equiv
      2\ln\frac{\LamD}{T}
    - 2\bigl(\ln(4\pi) - \gammaE\bigr)
  \,,&
  \Lf &\equiv \Lb + 4\ln2
  \,,\\
\label{eq:app:cconst}
  c &=
    \frac12\Bigl(
      \ln\frac{8\pi}{9}
    + (\ln\zeta_2)'
    - 2\gammaE
    \Bigr)
  \,,
\end{align}
where $\gammaE$ is the Euler-Mascheroni constant,
$\zeta_s=\zeta(s)$ the Riemann zeta function, and
$(\ln\zeta_s)'=\zeta'(s)/\zeta(s)$.
The 3D scalar thermal mass, quartic, portal, and temporal-scalar
couplings are
\begin{align}
\label{eq:app:mu3:2l}
  \mu_{3}^{2} &=
      \frac{T^2}{12}\bigl( 4\lambda_{\phi} + 3\gX^{2} \bigr)
    - \frac{T^{2}}{(4\pi)^{2}}\frac{\gX^{2}}{9}\bigl(
        2\gX^{2} - 6\lambda_{\phi}
      \bigr)
    - \frac{T^{2}}{(4\pi)^{2}}\Lb\Bigl(
        \frac{13}{12}\gX^{4}
      - 2\gX^{2}\lambda_{\phi}
      + \frac{10}{3}\lambda_{\phi}^{2}
      \Bigr)
    \nn[1mm] &\quad
    - \clog\frac{
        4\gXt^{4}
      - 8\gXt^{2}\lphit
      + 8\lphit^{2}
      + h_{3}^{2}
    }{(4\pi)^{2}}
  \,,\\[2mm]
\label{eq:app:lambda3}
  \lphit &=
    T\Bigl[
      \lambda_{\phi}
    + \frac{
        (2 - 3\Lb)\gX^{4}
      + \Lb\bigl( 6\gX^{2}\lambda_{\phi} - 10\lambda_{\phi}^2 \bigr)
      }{(4\pi)^2}
    \Bigr]
  \,,\\[2mm]
\label{eq:app:h3}
  h_{3} &=
    2\gX^{2} T\Bigl[
      1 - \frac{1}{(4\pi)^2}\Bigl(
          \frac{\Lb-4}{3}\gX^{2} - 8\lambda_{\phi}
        \Bigr)
    \Bigr]
  \,,\\[2mm]
\label{eq:app:kappa3}
  \kappa_{3} &=
    16\,\frac{\gX^{4}T}{(4\pi)^{2}}
  \,.
\end{align}
After integrating out the gauge modes
$\mXt \sim \gXt\,\phibg_3$ and
$m_{\rmii{$X_0$}} \sim \gXt\,T$ simultaneously,
the LO nucleation potential of eq.~\eqref{eq:Seff:LO}
takes the form
\begin{align}
\label{eq:app:VeffLO:U1}
  \Veff^\rmii{LO} &=
      \frac12 \mu_{3}^2 \phibg_3^2
    + \frac14 \lphit^{ } \phibg_3^4
    - \frac{1}{12\pi}\bigl(
        2 \mXt^3 + m_{\rmii{$X_0$}}^3
      \bigr)
    + \frac{1}{12\pi}\mD^3
  \,,
\end{align}
with
the last term cancelling the field-independent part of the potential~\cite{Kierkla:2023von} and
the 3D mass eigenvalues
\begin{align}
\label{eq:app:masses:U1}
  m_{\rmii{$H$}}^{2} &= \mu_{3}^{2} + 3\lphit\phibg_3^2
  \,,&
  m_{\rmii{$G$}}^{2} &= \mu_{3}^{2} + \lphit\phibg_3^2
  \,,\nn
  \mXt^{2} &= \gXt^2 \phibg_3^2
  \,,&
  m_{\rmii{$X_0$}}^{2} &= \mD^2 + \frac12 h_{3}\phibg_3^2
  \,.
\end{align}

\subsection{Next-to-leading order}
\label{sec:app:matching:nlo}

At NLO,
the soft expansion supplies a two-loop correction to the
effective potential and an effective kinetic operator on the LO bounce
$\phib$~\cite{Kierkla:2023von,Gould:2023ovu},
\begin{align}
\label{eq:app:Seff:NLO}
  S_3^\rmii{NLO}[\phib] =
    \int_\vec{x}\Bigl[
      \frac12 Z_{\phi_3}(\phib)\,(\partial_i\phib)^2
    + \Veff^\rmii{NLO}(\phib)
    \Bigr]
  \,,
\end{align}
where $\int_\vec{x} \equiv \int d^3\vec{x}$.
The NLO contributions are
suppressed relative to LO by the soft expansion parameter
$\sim\gX/\pi$.
For the $\UOneX$ model,
the NLO potential consists of temporal and
vector contributions,
\begin{align}
\label{eq:app:VeffNLO:U1}
  \Veff^\rmii{NLO} &=
  - \frac{1}{(4\pi)^2}\biggl[
      h_3\,\frac{m_{\rmii{$X_0$}}^{2} - \mD^2}{4}\Bigl(
        1 + \ln\frac{\Lamd^2}{4 m_{\rmii{$X_0$}}^{2}}
      \Bigr)
    + \gXt^{2} m_{\rmii{$X$}}^{2}\Bigl(
        1 + \ln\frac{\Lamd^2}{4 m_{\rmii{$X$}}^{2}}
      \Bigr)
    - \frac{\kappa_{3}}{8} m_{\rmii{$X_0$}}^{2}
    \biggr]
  \,,
\end{align}
where $\Lamd$ is the 3D EFT renormalization scale
defined in eq.~\eqref{eq:LamD:ref}, and
field-independent contributions are subtracted as in~\cite{Kierkla:2023von}.
The field normalization (kinetic term) for the nucleating scalar reads
\begin{align}
\label{eq:app:Zphi3:U1}
  Z_{\phibg_{3}} &=
    \frac{1}{48\pi}\biggl[
    - 22\,\frac{\gXt}{\phibg_{3}}
    + \frac14\,\frac{h_{3}^{2}\phibg_{3}^{2}}{m_{\rmii{$X_0$}}^{3}}
    \biggr]
  \,.
\end{align}
This NLO action defines
the \NLOdet{} contribution in eq.~\eqref{eq:Gamma:NLOdet}.
The \Daisy{} approximation instead corresponds to
the leading-order 3D EFT of
eq.~\eqref{eq:Seff:LO}
with effective parameters obtained from lower-order matching.

%
\section{Renormalization-group running and input fixing}
\label{sec:app:RG}

\subsection{One-loop $\beta$-functions}
\label{sec:app:RG:beta}

The 4D \MSbar{} couplings of the $\UOneX$ model run with the
renormalization scale $\LamD$ according to their one-loop
$\beta$-functions.
The \MSbar{} scale $\LamD$ is related to the scale $\mu$ of
dimensional regularization through
$\ln\LamD^2 \equiv \ln \mu^2 + \ln(4\pi) - \gammaE$.
Introducing the logarithmic running variable
\begin{align}
\label{eq:app:rge:param}
  t \equiv \ln \frac{\LamD^2}{\LamDin^2}
  \,,
\end{align}
with the input scale $\LamDin$,
the gauge and scalar couplings evolve at one loop
as~\cite{Hirvonen:2021zej}
\begin{align}
\label{eq:app:beta:gd}
  \partial_{t}\,\gX^2 &=
    \frac{1}{(4\pi)^2}\,\frac{\gX^{4}}{3}
  \,,\\
\label{eq:app:beta:lambda}
  \partial_{t}\,\lambda_{\phi} &=
    \frac{1}{(4\pi)^2}\Bigl(
        10\lambda_{\phi}^{2}
      - 6\gX^{2}\lambda_{\phi}^{ }
      + 3\gX^{4}
    \Bigr)
  \,.
\end{align}
These equations evolve the couplings between the input scale and the
matching scale entering the matching
relations~\eqref{eq:app:gX3}--\eqref{eq:app:kappa3}.

\subsection{Fixing the input parameters}
\label{sec:app:RG:inputs}

In the 3D~EFT we use tree-level vacuum values for the couplings;
this can be extended to one-loop corrected relations following
appendix~A of~\cite{Niemi:2021qvp} and~\cite{Lewicki:2024xan,Biondini:2026uds}.
The parameters of the conformal model are fixed by the
scheme~\cite{Kierkla:2022odc}:
\begin{itemize}
  \item[\bf In:]
    the gauge coupling $\gX$ and
    physical mass $\MX$
    (identified with the tree-level mass evaluated at the minimum)
    defined at the input scale $\LamDin = \MX$;
  \item[(i)]
    minimize the loop-level scalar potential
    fixing the quartic $\lambda_{\phi}$;
  \item[(ii)]
    run the parameters from $\LamDin$
    to the electroweak scale and match
    the portal coupling $\lambda_p$ and
    the Higgs self-coupling $\lambda_h$
    to generate the electroweak vacuum via
    $\mu_\rmii{SM}^2 = \lp\,\langle\Phi\rangle^2$ and
    reproduce the correct Higgs mass;
  \item[(iii)]
    run to
    the reference matching scale $\LamDref = \LamX\pi T$
    of eq.~\eqref{eq:LamD:ref}
    using the one-loop
    $\beta$-functions~\eqref{eq:app:beta:gd}--\eqref{eq:app:beta:lambda};
    the constant $\LamX$ is varied to probe higher-order corrections and
    set according to eq.~\eqref{eq:LamD:ref};
  \item[\bf Out:]
    \MSbar{} parameters as functions of the physical inputs and
    $T$, and the 3D~EFT parameters.
\end{itemize}
Standard Model extensions with the new
$\mathrm{U}(1)_\rmii{$X$}$ symmetry have
an additional parameter $g_\rmi{mix}$,
quantifying kinetic mixing between the $\UOneX$ and $\UOneY$ gauge fields,
which we set to zero at the input scale, $g_\rmi{mix}(\LamDin=\MX)=0$, for simplicity.
In the absence of fermions charged under both
$\UOneX$ and $\UOneY$,
$g_\rmi{mix}$ then vanishes at all scales.
Hence, we naturally avoid experimental dark photon constraints from
kinetic mixing~\cite{%
  Fabbrichesi:2020wbt,Caputo:2021eaa,Chang:2016ntp,Fradette:2014sza,Berger:2016vxi,
  Sung:2019xie,DeRocco:2019njg}.

The minimization can be further improved with
the full one-loop vacuum renormalization,
relating pole masses to physical two-point functions and
respecting their momentum dependence~\cite{Biondini:2026uds,Sojka:2024btp}.

{\small
%
\bibliographystyle{utphys}
\bibliography{ref}

}
\end{document}